\documentstyle[epsfig]{aipproc}

\begin{document}
\noindent hep-ph/9706446 \hfill {Saclay--SPhT/T97--61}

\title{Parton Densities from Collider Data\thanks{Presented at 
the 5${}^{\rm th}$ International Workshop on Deep Inelastic Scattering
and QCD, Chicago, Illinois, April 14--18, 1997}}

\author{David A. Kosower}
\address{Service de Physique Th\'eorique\thanks{Laboratory of the
{\it Direction des Sciences de la Mati\`ere\/}
of the {\it Commissariat \`a l'Energie Atomique\/} of France.}\\
CEA --- Saclay\\
F--91191 Gif-sur-Yvette cedex\\
France}

\maketitle

\begin{abstract}
Collider data can play an important role in determining the parton distribution
functions of the nucleon.  I outline a formalism which makes it possible to
use next-to-leading order calculations in such an extraction, while minimizing 
the amount of numerical computation required.
\end{abstract}

\section*{Introduction}

The  search for new physics  underlying  the standard model requires a
precise  understanding of known  physics.   At hadron colliders,  this
translates into   the  requirement  for  a   precise  understanding of
perturbative QCD, along with knowledge of the non-perturbative inputs:
$\alpha_s$ and  the parton distribution functions of  the nucleon.  In
recent years, our  detailed knowledge of perturbative  QCD  --- in the
form  of    matrix elements\cite{ME},  techniques   for combining
virtual   and      real   corrections     in    next-to-leading  order
calculations\cite{GG,GGK,FKS,CS} ---  has grown  tremendously.  The
same is true  of our knowledge  of $\alpha_s$\cite{alphas}, and of the
parton densities at small $x$\cite{Roberts}.

On the  other hand, issues which  might have  been thought settled ---
the parton distributions, and in particular the gluon distribution, at
moderate  $x$ --- have become unsettled.    Investigations by the CTEQ
group\cite{CTEQ-HJ}, in  light of the  supposed excess  of large-$E_T$
jets  claimed by CDF\cite{CDF-HJ},   have shown that deeply  inelastic
scattering   does {\it not\/}  constrain the   gluon distribution well
enough to allow a claim of new physics to be made  on the basis of the
inclusive jet distribution.  (The claim  of an excess is not, however,
supported by the D\O\ data\cite{D0}.)  More recently, CTEQ has pointed
out\cite{CTEQ-HERA}  that similar  statements   apply to the  high-$x$
valence quark  distributions  and  resulting   implications of    ZEUS
high-$Q^2$ data\cite{ZEUS}.

\def\yir{y_{\rm  IR}} To date,  the  parton densities  have been taken
solely as inputs in  calculations of collider processes.  To constrain
the gluon  distribution better,   however,  it would  be  desirable to
include  collider    data in     the    global fits.      It   is  the
triply-differential distribution\cite{Triply}     in     dijet events,
$d^3\sigma/dE_T d\eta_1 d\eta_2$,  that  will presumably give   us the
most useful information on the gluon distribution.  To obtain sensible
results using jet data, however, the use  of theory at next-to-leading
order is  a  bare minimum.  From  a  theoretical  point of  view, this
results  from the presence  of multiple  scales:  a jet  cross section
contains   not   only a  hard-scattering   scale,   but also  a  scale
characterizing  the `size' of the jet.   The perturbative expansion is
thus   not  merely   an  expansion  in    $\alpha_s$,  but  rather  in
$\alpha_s\ln^2 \yir$ and $\alpha_s \ln  \yir$, where $\yir$ is a ratio
of  scales, as  well.  In  order to  ascertain  that these potentially
dangerous  logarithms are under  control,   and that the  perturbative
calculation    is  reliable, we    must   calculate  (at    least)  at
next-to-leading order.  We must therefore perform fits to collider data
using NLO theoretical calculations.

\section*{Fitting to Data}

The parton densities depend on both the momentum fraction $x$
at which they are evaluated, and on a factorization scale $Q^2$.
Their evolution with changes in $Q^2$ is governed by the 
Altarelli-Parisi equation\cite{AltarelliParisi}, 
\begin{equation}
Q^2 {\partial f(x,Q^2)\over \partial Q^2} = 
\int_0^1 dy\,dz\;\delta(x-y z)\,P(y,Q^2) f(z,Q^2)\,,
\label{eq:AP}
\end{equation}
whose kernel $P$ can be computed perturbatively.
We may thus take the non-perturbative input to the parton densities to
be their values at some fixed scale $Q_0^2$, and it is these values that
we must fit to data.  This is usually done by picking a parametrization of
the form\cite{MRS},
\begin{equation}
f_i(x,Q_0^2) = A_i x^{-\lambda_i} (1-x)^{\beta_i} (1+\epsilon_i \sqrt{x} + \delta_i x)\,.
\end{equation}

For the traditional observables in
deeply-inelastic scattering data (e.g. $F_2$), one has relatively simple analytic NLO
formul\ae\ in terms of the parameters $\{A_i,\lambda_i,\beta_i,\cdots\}$, and
one can iteratively find a best fit by adjusting parameters, re-evaluating the
theoretical prediction, and comparing with measurements.

For jet distributions, such an iterative procedure would be extremely slow using
existing NLO jet programs, since each iteration involves re-running the jet program,
a matter of several hours if not a day even with present-day workstations.  
A better way of organizing the computation would thus be desirable.   

 We may expect that a better organization should be possible, because most of
the calculation --- computation of the perturbative matrix elements,
applying cuts, clustering, and the jet algorithm --- doesn't involve the parton
densities in an essential way.  I now give an outline of such a re-organization.
It bears certain similiarities to a formalism proposed by Graudenz, Hampel, Vogt,
and Berger\cite{GHVB} for jet production in deeply-inelastic scattering, but
is fully general and free of certain limitations present in that approach.

\def\LIPS{{\rm LIPS}}
\def\kset{\{k_i\}}

I shall explain the formalism in the context of a toy problem, leading-order
glueball-glueball scattering in quarkless QCD.  The same formalism, with
an appropriate sprinkling of indices and division into the different contributions
that arise at NLO, carries over to NLO fitting of both DIS and hadron-hadron collider
differential cross sections.  The $n$-jet cross section in glueball-glueball scattering,
subject to experimental cuts, is given by
\begin{eqnarray}
\sigma_n &=& \int_0^1 \int_0^1 dx_1 dx_2\; 
  \int d\LIPS(x_1 k_G+x_2 k'_{G} \rightarrow \{k_i\}_{i=1}^n) \nonumber\\
\label{eq:Cross}
 & &\hskip 10mm\times \vphantom{\int}
     f_{g\leftarrow G}(x_1,\mu^2_F(\kset,x_{1,2}))
     f_{g\leftarrow G}(x_2,\mu^2_F(\kset,x_{1,2}))\,\\
 & &\hskip 20mm\times \vphantom{\int}
     \alpha_s^n(\mu^2_R(\kset,x_{1,2}))\hat\sigma(g g\rightarrow \kset)\;
     J_{n\leftarrow n}(\kset)\nonumber
\end{eqnarray}
where $\LIPS$ stands for the Lorentz-invariant phase-space measure, 
$f_{g\leftarrow G}$ is the gluon distribution inside the glueball, and
$\hat\sigma$ stands for the usual leading-order partonic
differential cross section with the running coupling $\alpha_s$ set to 1.
Note that the $k_i$ are implicitly dependent on $x_1$ and $x_2$ as well.
The renormalization and factorization scales $\mu_R$ and $\mu_F$
--- typically something
like a jet $E_T$ --- also depend on $x_{1,2}$ and the final-state 
momenta.  The jet algorithm is represented by $J_{n\leftarrow n}$, which evaluates
to $1$ if the original $n$-parton configuration yields $n$ jets satisfying 
the experimental cuts, and $0$ otherwise.  

It is easiest to write down the Mellin transform of the 
solutions to the evolution equation~(\ref{eq:AP}), using a universal evolution
operator\cite{FP},
\begin{equation}
f_{g\leftarrow G}^z(Q^2) = E^z(\alpha_s(Q^2),\alpha_0) 
                             f_{g\leftarrow G}^z(Q_0^2)\,,
\end{equation}
where $z$ is the conjugate variable to $x$.
An inverse Mellin transform gives us back the parton distribution itself,
\begin{equation}
f_{g\leftarrow G}(x,Q^2) = 
{1\over 2\pi i}\int_{C} dz\; x^{-z} 
E^z(\alpha_s(Q^2),\alpha_0)  f_{g\leftarrow G}^z(Q_0^2)\,,
\label{eq:InverseMT}
\end{equation}
where the contour $C$ lies to the right of all singularities in $f$.

All the parameters (except $\alpha_0$) that we wish to fit are
contained in $f_{g\leftarrow G}^z(Q_0^2)$.  This function is {\it independent\/}
of all integration variables except $z$, and thus can be pulled out of
the numerical integrations in eqn.~(\ref{eq:Cross}).  The remaining
$z_{1,2}$ contour integrals are to be performed during the fitting
procedure, but this reduces to just a double sum of the gluon
distribution function multiplied
by precomputed numerical coefficients, and is vastly less time-consuming
than the evaluation of the cross-section~(\ref{eq:Cross}).  That is, each
step of a fitting procedure requires computing,
\begin{equation}
-{1\over 4\pi^2} \int_{c-i\infty}^{c+i\infty} dz_1\,
\int_{c-i\infty}^{c+i\infty} dz_2\; 
f_{g\leftarrow G}^{z_1}(Q_0^2)
f_{g\leftarrow G}^{z_2}(Q_0^2) \Sigma^{z_1, z_2}\,,
\label{eq:FitStep}
\end{equation}
where $\Sigma^{z_1,z_2}$ are precomputed coefficients given by
\begin{eqnarray}
\Sigma^{z_1,z_2} &=& \int_0^1 \int_0^1 dx_1 dx_2\; 
  \int d\LIPS(x_1 k_G+x_2 k'_{G} \rightarrow \{k_i\}_{i=1}^n) x_1^{-z_1} x_2^{-z_2} 
\nonumber\\
& &\hskip 10mm\times
E^{z_1}(\alpha_s(\mu^2_F(\kset,x_{1,2})),\alpha_0)
E^{z_2}(\alpha_s(\mu^2_F(\kset,x_{1,2})),\alpha_0)\\
& &\hskip 13mm\times
     \alpha_s^n(\mu^2_R(\kset,x_{1,2}))\hat\sigma(g g\rightarrow \kset)\;
     J_{n\leftarrow n}(\kset)\,.\nonumber
\end{eqnarray}
We see that $\Sigma$ is a `cross section' computed with $x^{-z} E^z$ replacing
the usual parton distribution functions.

To fit the remaining   parameter, $\alpha_0$, the running coupling  at
$Q_0^2$,  we can   procede  as  follows.   Generate  the  coefficients
$\Sigma^{z_1,z_2}$  for  a set  of  $\alpha_0$ around  a ``canonical''
value  (e.g.  $\alpha_s(M_Z^2)   =  0.118$),   and  then  fit    using
interpolation.  While this approach would be vastly too time-consuming
for   a  large number  of  parameters, it  is   acceptable  for a lone
parameter.

As mentioned above, a similar decomposition works for differential
cross sections in full QCD.

The contour integrals in eqn.~(\ref{eq:FitStep}) must be performed
numerically.  We must first choose the contours, and then a set of points
along each contour; the integral is then approximated, as usual, by
a sum of the integrand evaluated at these points.  The contour in
eqn.~(\ref{eq:InverseMT}) can be deformed freely into the left half-plane
for $z\rightarrow\infty$, since all the singularities of the integrand are
along the real axis; we can choose an efficient quadratic contour using
the same ideas as described in ref.~\cite{Evolution} for evolving parton
distributions.

\end{document}